\documentclass[aps,prb,twocolumn,superscriptaddress]{revtex4}

\usepackage{amsmath}
\usepackage{epsf}
\usepackage{epsfig}

\begin{document}

\title{High-capacity hydrogen storage by metallized graphene}

\author{C. Ataca}
\affiliation{Department of Physics, Bilkent University, Ankara
06800, Turkey}\affiliation{UNAM-Institute of Materials Science and
Nanotechnology, Bilkent University, Ankara 06800, Turkey}
\author{E. Akt\"{u}rk}
\affiliation{UNAM-Institute of Materials Science
and Nanotechnology, Bilkent University, Ankara 06800, Turkey}
\author{S. Ciraci} \email{ciraci@fen.bilkent.edu.tr}
\affiliation{Department of Physics, Bilkent University, Ankara
06800, Turkey}
\affiliation{UNAM-Institute of Materials Science
and Nanotechnology, Bilkent University, Ankara 06800, Turkey}
\author{H. Ustunel}
\affiliation{Department of Physics, Middle East Technical
University, Ankara 06531, Turkey}

\begin{abstract}
First-principles plane wave calculations predict that Li can be
adsorbed on graphene forming a uniform and stable coverage on both
sides. A significant part of the electronic charge of the Li-$2s$
orbital is donated to graphene and is accommodated by its distorted
$\pi^*$-bands. As a result, semimetallic graphene and
semiconducting graphene ribbons change into good metals. It is
even more remarkable that Li covered graphene can serve as a
high-capacity hydrogen storage medium with each adsorbed Li
absorbing up to four H$_2$ molecules amounting to a gravimetric
density of 12.8 wt \%.
\end{abstract}

\maketitle

Developing safe and efficient hydrogen storage is essential for
hydrogen economy.\cite{coontz} Recently, much effort has been
devoted to engineer carbon based
nanostructures\cite{sefa,taner,heben,engin} which can absorb H$_2$
molecules with high storage capacity, but can release them easily
in the course of consumption in fuel cells. Insufficient storage
capacity, slow kinetics, poor reversibility and high
dehydrogenation temperatures have been the main difficulties
towards acceptable media for hydrogen storage.

Recently, graphene, a single atomic plane of graphite, has been
produced\cite{novo} showing unusual electronic and magnetic
properties. In this letter, we predict that metallized graphene
can be a potential high-capacity hydrogen storage medium. The
process is achieved in two steps: Initially, graphene is
metallized through charge donation by adsorbed Li atoms to its
$\pi^{*}$- bands. Subsequently, each positively charged Li ion can
absorb up to four H$_2$ by polarizing these molecules. At the end,
the storage capacity up to the gravimetric density of $g_d$=12.8
wt \% is attained. These results are important not only because
graphene is found to be a high capacity hydrogen storage medium,
but also because of its metallization through Li coverage is
predicted.

Our results have been obtained by performing first-principles
plane wave calculations using ultra-soft
pseudopotentials.\cite{vander} We used Local Density Approximation
(LDA), since the van der Waals contribution to the Li-graphene
interaction has been shown\cite{lda} to be better accounted by
LDA. Numerical results have been obtained by using
VASP,\cite{vasp} which were confirmed by using the PWSCF
code.\cite{pwscf}  A plane-wave basis set with kinetic energy
cutoff $\hbar^2 |\textbf{k}+\textbf{G}|^2/2m$ = 380 eV has been
used. In the self-consistent potential and total energy
calculations the Brillouin zone has been sampled by (19x19x1) and
(9x9x1) special mesh points in \textbf{k}-space for (2$\times$2)
and (4$\times$4) graphene cells, respectively. Atomic positions in
all structures are optimized using the conjugate gradient method.
Convergence is achieved when the difference of the total energies
of last two consecutive steps is less than $10^{-6}$ eV and the
maximum force allowed on each atom is less than $10^{-2}$ eV/\AA.
All configurations studied in this work have also been calculated
by using spin-polarized LDA, which were resulted in non-magnetic
ground state.

Adsorption of a single (isolated) Li atom on the hollow site of
graphene (i.e. H1-site above the center of hexagon) is modelled by
using (4$\times$4) cell of graphene with 1.70 \AA~ minimum
Li-graphene distance and with a minimum Li-Li distance of 9.77
\AA, resulting in a binding energy of $E_L$=1.93 eV. Upon
adsorption Li atom donates part of the charge of its $2s$ state to
the more electronegative carbon atoms at its proximity. Despite the
ambiguities in determining the atomic charge,  L\"{o}wdin analysis
estimates that Li becomes positively charged by donating $q \sim $0.35
electrons (but $q \sim  0.9$ electrons according to Bader
analysis\cite{bader}). The energy barrier to the diffusion of a
single Li atom on the graphene sheet through top (on-top of carbon atoms) and
bridge (above the carbon-carbon bond) sites
are calculated to be $\Delta Q$=0.35 eV and 0.14 eV, respectively.

\begin{center}
\begin{figure*}
\includegraphics[scale=0.7]{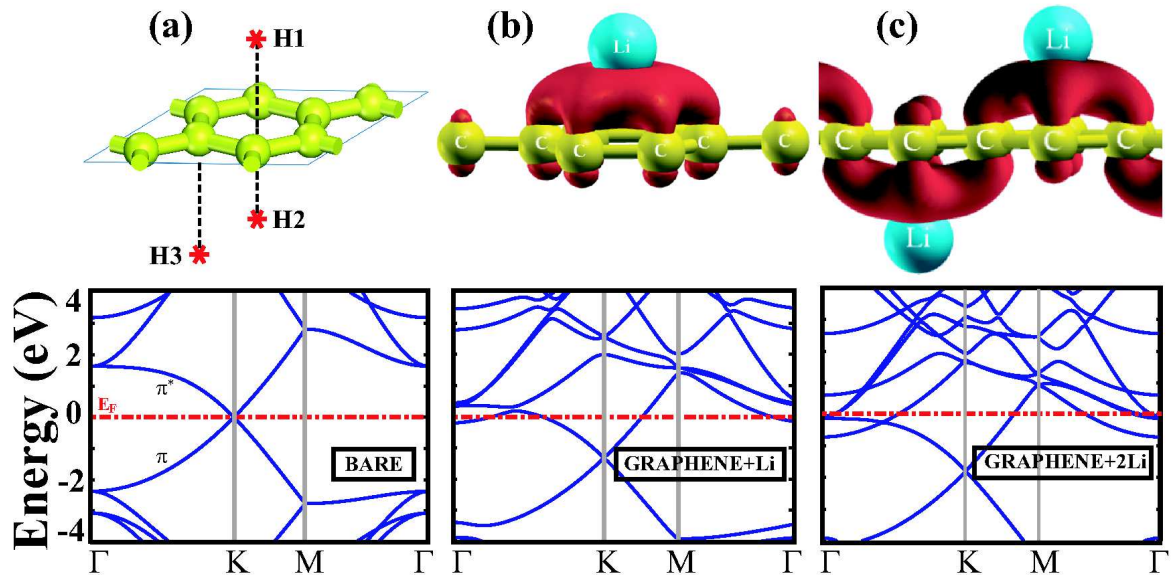}
\caption{(Color online) (a) Various adsorption sites H1, H2 and H3
on the (2$\times$2) cell (top panel) and energy band structure of
bare graphene folded to the (2x2) cell (bottom panel). (b) Charge
accumulation, $\Delta \rho^+$, calculated for one Li atom adsorbed
to a single side specified as H1 (top) and corresponding band
structure. (c) Same as (b) for one Li atom adsorbed to H1-site,
second Li adsorbed to H3-site of the (2$\times$2) cell of
graphene. Zero of band energy is set to the Fermi energy, $E_F$. }
\label{fig:band}
\end{figure*}
\end{center}

Lithium atoms can form a denser coverage on the graphene with a
smaller Li-Li distance of 4.92 \AA~ forming the (2$\times$2)
pattern. Owing to the repulsive interaction between positively
charged Li atoms, the binding energy of Li atom is smaller than
that of the (4$\times$4) cell. For H1 adsorption site (see
Fig.\ref{fig:band}(a)), the binding energy is calculated to be
$E_L$=0.86 eV. The binding energies are relatively smaller at the
bridge and top sites, and are 0.58 and 0.56 eV, respectively. The
binding energy of the second Li for the double sided adsorption
with H1+H2 and H1+H3 configurations described in
Fig.\ref{fig:band} (a), are $E_L$=0.82 and 0.84 eV, respectively.
The same binding energies for H1+H2, and H1+H3 geometries on the
(4$\times$4) cell are relatively larger due to reduced repulsive
Li-Li interaction, namely $E_L$=1.40 eV and 1.67 eV, respectively.
The coverage of Li on the (2$\times$2) cell is $\Theta$=12.5 \%
(i.e. one Li for every 8 carbon atoms) for H1 geometry and
$\Theta$=25 \% for either H1+H2 or H1+H3 geometries. Metallic
charge accumulated between Li and graphene weakens the interaction
between Li atoms which are adsorbed at different sites of
graphene. Further increasing one-sided coverage of Li to
$\Theta$=25\%  with H1 geometry (or two-sided coverage to 50 \%
with H1+H2 or H1+H3 geometries) appears to be impossible due to
strong Coulomb repulsion between adsorbed Li ions and results in a
negative binding energy ($E_L \sim$ - 2.5 eV). On the other hand,
the total binding energy of all Li atoms adsorbed on a(2$\times$2)
cell with the H1+H2 (H1+H3) geometry corresponding to $\Theta$=25
\% is 3.23 eV (3.12 eV) higher than that of Li atom adsorbed on
the (4$\times$4) cell with the same geometry corresponding to
$\Theta$=6.25 \%. Hence, since the cluster formation is hindered
by the repulsive interaction between the adsorbed ions, a stable
and uniform Li coverage on both sides of graphene up to $\Theta =
$ 25 \% can be attained.

\begin{center}
\begin{figure*}
\includegraphics[scale=0.7]{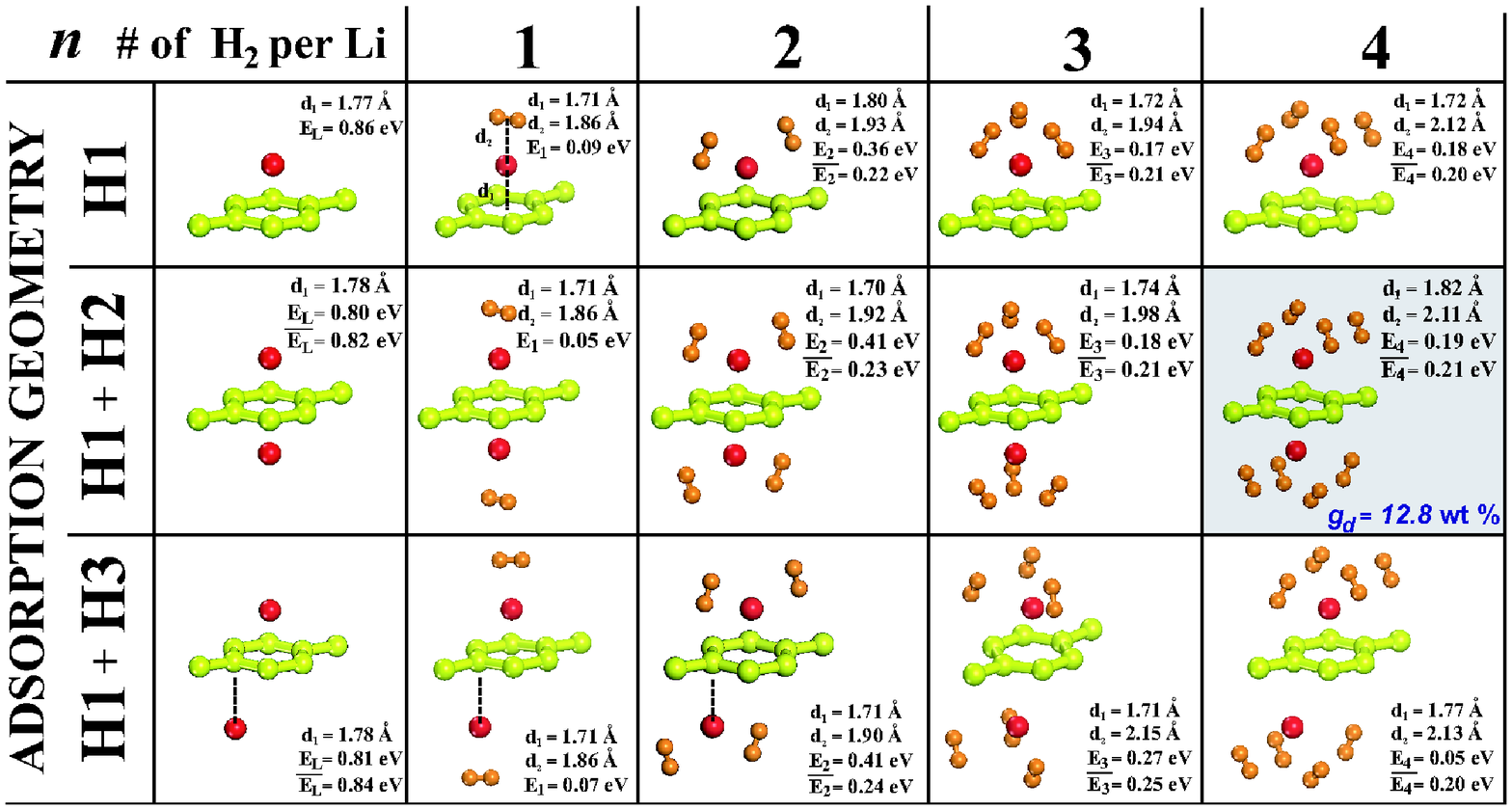}
\caption{(Color online) Adsorption sites and energetics of Li
adsorbed to the (2$\times$2) cell of graphene and absorption of
H$_2$ molecules by Li atoms. $E_L$ is the binding energy of Li
atom adsorbed to H1-site, which is a minimum energy site. For
H1+H2 or H1+H3 configuration corresponding to double sided
adsorption, $E_L$ is the binding energy of second Li atom and
$\overline E_L$ is the average binding energy. For H1, H1+H2 and
H1+H3 configurations, $E_1$ is the binding energy of the first
H$_2$ absorbed by each Li atom; $E_n$ ($n$=2-4) is the binding
energy of the last $n^{th}$ H$_2$ molecule absorbed by each Li
atom; $\overline E_n$ is the average binding energy of $n$ H$_2$
molecules absorbed by a Li atom. Shaded panel indicates the most
favorable H$_2$ absorbtion configuration.}\label{fig:energy}
\end{figure*}
\end{center}

The charge accumulation and band structure calculated for the H1
and H1+H3 adsorption geometries are presented in
Fig.\ref{fig:band} (b) and (c), respectively. Isosurface plots of
charge accumulation obtained by subtracting charge densities of Li
and bare graphene from that of Li which is adsorbed to graphene,
$\Delta \rho^{+}$, display positive values. As a result of Li
adsorption, the charge donated by Li is accumulated between Li and
graphene and is accommodated by $2p\pi^*$-bonds of carbons. The
empty $\pi^*$- bands become occupied and eventually get distorted.
Occupation of distorted graphene-$\pi^*$ bands gives rise to the
metallization of semimetallic graphene sheets. By controlled Li
coverage one can monitor the position of Fermi energy in the
linear region of bands crossing at the K-point of the Brillouin
zone. Metallization is also important for zigzag and armchair
graphene nanoribbons, since both are semiconductors with their
energy gaps depending strongly on the widths of these
ribbons.\cite{cohen} Segments of these ribbons metallized by Li
adsorption may be interesting for their electronic and spintronic
applications. For example, a junction of two nanoribbons with and
without Li adsorbed segments can serve as a Schottky barrier.

Sodium, a heavier alkali metal, can be bound to graphene with
$E_b$=1.09 eV at H1 site. However, the energy difference between
top, bridge and H1 sites are minute due to relatively larger
radius of Na. Upon adsorption, graphene and graphene nanoribbon
are metallized. Nevertheless, Na is not suitable for hydrogen
storage because of its heavier mass and very weak binding to H$_2$
molecules. Two dimensional BN-honeycomb structure, being as a
possible alternative to graphene, has very weak binding to Li
($\sim$ 0.13 eV) and hence it is not suitable for hydrogen
storage.

The absorption of H$_2$ molecules by Li + graphene in H1, H1+H2
and H1+H3 geometries. A summary of our results about the H$_2$
absorption are presented in Fig. \ref{fig:energy}. The binding
energy of the first absorbed H$_2$, which prefers to be parallel
to graphene, is generally small. However, when two or more H$_2$
molecules are absorbed by the same Li atom, the binding geometry
and mechanism change and result in a relatively higher binding
energy. All H$_2$ molecules are tilted so that one of two H atoms
of each absorbed H$_2$ molecules becomes relatively closer to the
Li atom. A weak ionic bond forms through a small amount change
($\gtrsim$ 0.1 electrons) transferred from Li and graphene to
nearest H atoms of absorbed H$_2$ molecules. At the end, H atoms
receiving charge from Li becomes negatively charged and the
covalent H$_2$ bond becomes polarized. Weak ionic bond, attractive
Coulomb interaction between positively charged Li and negatively
charged H and weak van der Waals interaction are responsible for
the formation of mixed weak bonding between H$_2$ molecules and
Li+graphene complex. Here the bonding interaction is different
from the Dewar-Kubas interaction \cite{dewar} found in
H$_2$-Ti+C$_{60}$ or carbon nanotube complexes.\cite{taner} As the
number of absorbed H$_2$, $n$, increases, the positive charge on
Li as well as the minimum distance between H$_2$ and Li slightly
increases. No matter what the initial geometry of absorbed H$_2$
molecules would be, they are relaxed to the same final geometry
presented in Fig.\ref{fig:energy} for any given $n$. We found no
energy barrier for a H$_2$ molecule approaching the absorbed H$_2$
when $n \le 4$. Note that the dissociative absorption of H$_2$
molecules do not occur in the present system. The energy barrier
for the dissociation of H$_2$ near Li to form Li-H bond is $\sim
2$ eV.\cite{zhao} Moreover, dissociation of H$_2$ to form two C-H
bonds at the graphene surface is energetically unfavorable by 0.7
eV.

Maximum number of absorbed H$_2$ per Li atom is four, and the
maximum gravimetric density corresponding to H1+H2 geometry at
$\Theta$=25 \% coverage is $g_d$=12.8 wt \%. This is much higher
than the limit  ($g_d$=6 wt \%) set for the feasible H$_2$ storage
capacity. Note that only for $n=4$, H1+H2 absorption geometry has
slightly lower energy than H1+H3 geometry. This is a remarkable
result indicating another application of graphene as a high
capacity storage medium. Here Li+graphene complex is superior to
Ti+C$_{60}$ or carbon nanotube complexes since Li is lighter. Even
though graphene by itself is stable\cite{geim,katnelson}, more
stable form is obtained by Li adsorption on graphene due to strong
Coulomb repulsion between adsorbed Li atoms. Moreover, Li covered
graphene is resistant to clustering of adsorbed Li atoms. Earlier,
Durgun et al.\cite{engin} has predicted that ethylen+Ti complex
can store H$_2$ up to $g_d$=14.4 wt \% per molecule. Later, their
results have been confirmed experimentally.\cite{shiravam} We
believe that hydrogen storage by the Li covered graphene is
interesting, since it may not require encapsuling and hence may
yield even higher effective $g_d$.

In conclusion, two crucial features of Li covered graphene
revealed in this paper may be of technological interest. These are
high metallicity and high hydrogen storage capacity of graphene
functionalized by Li atoms. Graphene nanoribbons  metallized
through adsorbed Li atoms can be used as interconnects between
graphene based spintronic devices. Graphene functionalized by Li
can also serve as a medium of hydrogen storage. As far as
efficiency in storage is concerned, graphene may be superior to
carbon nanotubes, because its both sides are readily utilized.
Cell configurations formed by different junctions of
graphene\cite{kawai} functionalized by Li atoms are expected to
yield higher surface/volume ratio and hence to provide efficient
H$_2$ storage in real applications.

This research was supported by the Scientific and Technological
Research Council of Turkey under the project TBAG 104536. Part of
computational resources have been provided through a grant
(20242007) by the National Center for High Performance Computing,
Istanbul Technical University.

\end{document}